\documentclass[a4paper,10pt,prl,twocolumn]{revtex4-1}
\usepackage{amssymb,amsfonts,amsmath,graphicx,times,times,amssymb}
\usepackage{soul}
\usepackage{hyperref}

\date{\today}

\begin{document}

\author{David Edward Bruschi}
\affiliation{York Centre for Quantum Technologies, Department of Physics, University of York, YO10 5DD Heslington, UK}
\email{david.edward.bruschi@gmail.com}

\title{On the gravitational nature of energy}

\begin{abstract}
We propose the idea that not all energy is a source of gravity. We discuss the role of energy in the theory of gravitation and provide a formulation of gravity which takes into account the quantum nature of the source. We show that gravity depends dramatically on the entanglement present between the constituents of the Universe.
Applications of the theory and open questions are also discussed. 
\end{abstract}

\maketitle


\textit{Does all energy have a weight?} 
A century has passed since the original proposal that energy is the source of gravity. The gravitational nature of energy is now considered a pillar of physics. A change in this view would bring about dramatic consequences to our understanding of the laws of Nature.
 
In this work we propose the idea that not all energy gravitates. This idea stems from considering the role and nature of energy in quantum field theory \cite{Srednicki:2007} and semiclassical gravity \cite{BandD,Flanagan:Eanna:1996}, and by taking into account recent advances in the field of quantum thermodynamics \cite{Goold:Huber:2016}. 
On the one hand, quantum field theory in curved spacetime and semiclassical gravity cannot account for the quantum nature of gravity, nor can they account for the effects induced by the quantum nature of fields on the gravitational field itself, a phenomenon known as \textit{backreaction} \cite{Flanagan:Eanna:1996}. On the other, quantum thermodynamics extends the concepts of work, energy and entropy to the quantum realm, far from the thermodynamical limit where statistical fluctuations around the mean become significant \cite{Goold:Huber:2016}. 
Furthermore, another important difference arises between the two main theories or Nature available, i.e., quantum mechanics and general relativity. Gravity is an intrinsically local theory, while quantum mechanics, on the other hand, is a theory with nonlocal features, i.e., entanglement. A prospective theory of gravitating quantum matter must be able to reconcile these two seemingly incompatible features.

We propose a novel formulation of the field equations of gravity that introduces genuine quantum features as a contribution to the source of the dynamics of spacetime. We discuss the main features of the theory and compare preliminary predictions with known results from quantum field theory in curved spacetime \cite{BandD}. We find that the contribution to gravity of excited fields, i.e., associated with the presence of matter and energy, dramatically depends on their quantum state. For example, quantum matter in a thermal state does contain excitations (particles) but might not contribute to the gravitational field at all. 
In this sense, gravity depends dramatically on the interplay between the purity of the state of the Universe and the purity of the reduced states of its constituents. 

In this work we consider a classical theory of gravity, with the metric $g_{\mu\,\nu}$, in a (3+1)-dimensional spacetime \cite{Misner:Thorne:1973}. In standard approaches, Einstein's equations place classical matter and fields with stress energy tensor $T_{\mu\,\nu}$ as the source of gravity. They read
\begin{align} \label{einstein:equations}
G_{\mu \nu}=\frac{8\,\pi\,G_N}{c^4}\,T_{\mu \nu},
\end{align}
where $G_N$ is Newton's constant and $G_{\mu \nu}:=R_{\mu \nu}-\frac{1}{2}\,g_{\mu \nu}\,R$ is Einstein's tensor. The Ricci tensor $R_{\mu \nu}$ contains derivatives of the metric, and $R$ is its trace, i.e. $R:=g^{\mu \nu}\,R_{\mu \nu}$.

The theory assumes that the source $T_{\mu \nu}$ does not interact with any form of quantum matter (i.e., quantum fields) that propagates on the classical background. Einstein's theory of gravity has been extensively studied in the past and we refer to standard literature for details \cite{Misner:Thorne:1973}. 

Particles can be modelled as excitations of a quantum field $\phi$ that propagates on the classical generally curved background, or spacetime \cite{BandD}. For simplicity, a particle is an excitation of a massive scalar field $\phi(x^{\mu})$ that obeys the Klein-Gordon equation of motion $(\square+m^2)\phi=0$. The d'Alambertian operator $\square$ for curved spacetime is defined by $\square \equiv (\sqrt{-g})^{-1}\, \partial_{\mu}\, [\sqrt{-g}\, g^{\mu \nu}\, \partial_{\nu}]$. 
In this formalism, one encounters the problem of vacuum divergences, i.e., the vacuum state expectation value of many physically relevant quantities, such as $\langle0|T_{\mu \nu}|0\rangle$, is divergent \cite{Srednicki:2007}. These issues can be effectively resolved in flat spacetime, and partially in curved spacetime, by employing renormalisation techniques \cite{BandD}.

The Hawking effect \cite{Hawking:1974} and the Unruh effect \cite{Unruh:1976} are among the most exciting and well known predictions of quantum field theory in flat and curved spacetime. The former predicts that a black hole of mass $M$ radiates particles as a black body at temperature $T_H$ inversely proportional to the mass $M$. Analogously, the latter predicts that a uniformly accelerating observer, with constant proper acceleration $a$ and equipped with a particle detector, will detect particle excitations with a probability that is thermally distributed at temperature $T_U$ directly proportional to the proper acceleration $a$.  
A standard way to understand the basic features of these effects is to consider particle emission processes in Schwarzschild and Rindler spacetimes \cite{BandD}. Both cases are very special, since the spacetime is static. In this case the effects are independent of time, i.e., at any time, the hovering observer is immersed in the thermal bath of particles and the detector of the accelerated observer keeps ticking with the same probability per unit time. This poses interesting challenges, such as how to reconcile the energy carried by the detected excitations and their role in the gravitational process.

We highlight the fact that, in both the scenarios, there exist two (or more) different vacua of the theory given the high degree of symmetry and time translation invariance \cite{BandD}. In both cases, the vacua considered, say $|0\rangle$ and $|0'\rangle$, are not arbitrarily related \cite{BandD}. For our purposes, it is enough to recall that the two vacua are related by a ``unitary'' Bogoliubov transformation \cite{BandD}. The existence of these different vacua, and their role in explaining these renown effects, is paramount to our work.

The key point we want to emphasize is that, in both setups, energy is somehow associated with the effect and it is not clear how the energy is provided and where this energy is stored. Furthermore, since energy is the source of gravity, this leaves us with the open question of how to reconcile the existence of the (possibly infinite) energy necessary to witness the effects with the very theory of gravity that defines the arena where this energy is present in the first place. Therefore, we find it natural to ask the following question: \textit{why doesn't this energy, or this ensemble of particles, gravitate}? 

We believe that the problem stems from the nature and notion of energy. In classical physics, energy is a well defined concept. Other forms of energy, such as work or entropy, are not fundamental. Entropy, in particular, measures the amount of ignorance we have of a classical system, and the amount of ``waste'' that will be produced in a given task by not being able to access and employ all of the available information. In principle, the state of classical systems can be completely defined and all degrees of freedom measured with infinite precision. On the other hand, quantum mechanical systems don't enjoy these properties. In particular, a single value of energy is meaningless \cite{Goold:Huber:2016}, and entropy becomes something more fundamental, since it is intimately related to a global property of a quantum state, \textit{entanglement} \cite{Einstein:Podolsky:1935}. This genuine quantum feature implies an irreducible entropy of the reduced system \cite{Goold:Huber:2016}. 

Recent developments in the field of quantum thermodynamics allow us to extend notions of entropy and energy to single quantum systems \cite{Goold:Huber:2016}. More interestingly, it has been shown that not all energy can be employed for work, and this depends on the entropy of the system \cite{Goold:Huber:2016}. In particular, it is possible to extract all energy from a pure state, \textit{except} for the energy stored in the ground state of the system. Conversely, less energy can be extracted by mixed systems, which are plagued with intrinsic entropy \cite{Goold:Huber:2016}. These observations stimulated recent work aimed at understanding how entanglement contributes to the gravitational field \cite{Bruschi:2016}.

In light of all of these considerations, and motivated by the reasonings above, we propose the idea that \textit{not all energy is a source of gravity}. We suggest that only the energy that can be extracted from a system and converted into work will contribute to the gravitational field. This radical departure from the standard theory of gravity allows us to introduce a nonlocal element inherent to quantum mechanics, i.e., the entropy of entanglement, into a fundamentally local theory, i.e., gravity. The source of gravity becomes the \textit{extractible work} \cite{Allahverdyan:Balian:2004}, and we propose a field equation of the form
\begin{align} \label{quantum:einstein:equations}
G_{\mu \nu}=\frac{8\,\pi\,G_N}{c^4}\,\left[\langle \hat{T}_{\mu \nu}\rangle_{\rho}-\langle \hat{T}_{\mu \nu}\rangle_{\rho_{p.}}\right],
\end{align}
where the state $\rho$ is the initial state considered and the state $\rho_{p.}$ is the (unique) passive state that has the same spectrum to $\rho$, i.e., is obtained by $\rho$ through a unitary operation \cite{Horodecki:Oppenheim:2013}. The reason why we restrict to unitary operations is that the universe is believed to be a closed system. The source of gravity in \eqref{quantum:einstein:equations} is formally analogous to the ergotropy, which quantifies the extractible work in close systems, i.e., systems that evolve unitarily \cite{Horodecki:Oppenheim:2013}. When we talk about extractible work, we conisder the maximal extractible work from all observes (with local operations). More specifically, within standard quantum mechanics, the ergotropy is defined by the quantity $\langle \hat{H}\rangle_{\rho}-\langle \hat{H}\rangle_{\rho_{p.}}$, where $\hat{H}$ is the Hamiltonian of the system \cite{Allahverdyan:Balian:2004}. In scenarios with timelike Killing vectors, equivalently with a preferred notion of time \cite{BandD}, one can meaningfully recover the Hamiltonian through $\int d\Sigma\,\hat{T}_{\tau \tau}=\hat{H}$, where $\Sigma$ is an appropriate three-dimensional Cauchy hypersurface and $\partial_{\tau}$ is the orthogonal Killing vector field \cite{BandD} that defines the direction of time \cite{Srednicki:2007}. This motivates our choice of the expression $\langle \hat{T}_{\mu \nu}\rangle_{\rho}-\langle \hat{T}_{\mu \nu}\rangle_{\rho_{p.}}$ as the source of gravity. We note that the passive state $\rho_{p.}$ has the role of a ``reference'' state, since it is not possible to extract any work from a passive state \cite{Pusz:Woronowicz:1978}. Instructively, thermal states are an example of passive states. 
We conclude these remarks by adding that the gravitational part of the theory, i.e. the metric, remains classical. In this sense, we have not proposed a quantum theory of gravity.

The field equations \eqref{quantum:einstein:equations} depend on the metric, which in turn depends on the quantum nature of the source itself. In order to be consistent with the formulation of equation \eqref{quantum:einstein:equations}, we propose that expectations values of any relevant operator need to be taken in the same fashion as in the right hand side of \eqref{quantum:einstein:equations}, i.e., as a difference between an average over the state $\rho$ and an average on the corresponding passive state $\rho_{p.}$. We will comment on this at the end of this work.

The main equation \eqref{quantum:einstein:equations} tells us that the \textit{extractible}, or employable, fraction of energy is relevant, not the absolute content of it. The so called ``classical limit'' can be recovered when the fields considered are highly excited, or in a ``classical state''. This will correspond to the scenarios where $\langle \hat{T}_{\mu \nu}\rangle_{\rho}-\langle \hat{T}_{\mu \nu}\rangle_{\rho_{p.}}\sim T_{\mu \nu}$, and $T_{\mu \nu}$ is the classical stress-energy tensor, with good approximation. As an example, relativistic fluids are well understood \cite{Endlich:Nicolis:2011}, and could in principle be quantised. The classical behaviour would correspond to the standard stress energy tensor $T_{\mu \nu}=(p+\rho)\,u_{\mu}\,u_{\nu}+p\,g_{\mu \nu}$ for fluids. In this way we recover Einstein's theory of  gravity. In the classical limit, (almost) all energy can be potentially extracted or, equivalently, the vacuum state of a classical system contains (almost) no energy.

If the state of the universe is pure, the corresponding  passive state $\rho_{p.}$ is the vacuum of the theory. In this case $\langle \hat{T}_{\mu \nu}\rangle_{\rho_{p.}}=\langle0| \hat{T}_{\mu \nu}|0\rangle$ and equation \eqref{quantum:einstein:equations} is ``renormalised'' by the vacuum contribution. We emphasise that this is the vacuum of the theory, \textit{not} the vacuum of the test fields that propagate. However, we recall that in quantum field theory scenarios there might be more than one vacuum of the theory. Therefore we proceed to analyse the role of different vacua in our theory. Different vacua can be related to different quantisation procedures \cite{BandD}. The quantisation procedures are not arbitrary, in general, but are related to each other by means of a Bogoliubov transformation \cite{BandD}. This class of transformations is paramount in quantum field theory in curved spacetime and, as mentioned before, is at the core of celebrated effects, such as the Hawking effects \cite{Hawking:1974}, the Unruh effect \cite{Unruh:1976} and particle creation in an expanding universe \cite{BandD}. 
Let us assume that we pick a quantisation procedure, with vacuum $|0\rangle$, and employ the initial state $\rho=|0\rangle\langle0|$ with corresponding passive state $\rho_{p.}=\rho$. In this case, our equation  \eqref{quantum:einstein:equations} gives $G_{\mu \nu}=0$ which consistently provides a flat spacetime to a source in its vacuum state. However, if there is a different choice of vacuum state $|0'\rangle$, we then start from the vacuum state $\rho'=|0'\rangle\langle0'|$ and obtain, again $G_{\mu' \nu'}=0$, since the corresponding passive state is $\rho'_{p.}=|0'\rangle\langle0'|$, and \textit{not} the state $\rho_{p.}=|0\rangle\langle0|$. The conclusion we draw from this is that different vacua of the theory \textit{always} provide Minkowski spacetime. If a detector located in this spacetime registers particle counts, this will occur because of interactions with other agents, not because of any energy extraction process from the vacuum.

We now proceed to discuss the role and the interplay of the purity of the initial state, the purity of any reduced state of the system and the presence of an interacting theory of fields. We start by assuming that the global state $\rho$ of the universe is pure, i.e., $\rho^2=\rho$. This reflects the fact that the initial state considered contains all the information to be fully characterised at any time. We then assume we can separate the fields (or matter) into bulk B and the system S, where the ``system'' consists of all fields of direct interest to us. In a non interacting theory, the stress energy tensor reads $\hat{T}_{\mu \nu}=\hat{T}_{\mu \nu}^{B}+\hat{T}_{\mu \nu}^{S}$. Our main equation \eqref{quantum:einstein:equations} now reads
\begin{align} \label{quantum:einstein:equations:explicit}
G_{\mu \nu}=&\frac{8\,\pi\,G_N}{c^4}\,\left[\langle \hat{T}_{\mu \nu}^{B}\rangle_{\rho}-\langle0| \hat{T}_{\mu \nu}^{B}|0\rangle+\langle \hat{T}_{\mu \nu}^{S}\rangle_{\rho}-\langle0| \hat{T}_{\mu \nu}^{S}|0\rangle\right],
\end{align}
where we have used the fact that the global state is pure to write $\rho_{p.}=|0\rangle\langle0|$, and $|0\rangle$ is the vacuum of the full theory corresponding to the state $\rho$.

If the state $\rho$ is separable in the bulk-field bipartition, i.e., $\rho=\rho_{B}\otimes\rho_{S}$, we obtain $\langle \hat{T}_{\mu \nu}\rangle_{\rho}-\langle \hat{T}_{\mu \nu}\rangle_{\rho_{p.}}=\langle \hat{T}^{B}_{\mu \nu}\rangle_{\rho_{B}}-\langle0_b| \hat{T}^{B}_{\mu \nu}|0_B\rangle+\langle \hat{T}^{S}_{\mu \nu}\rangle_{\rho_{S}}-\langle0_S| \hat{T}^{S}_{\mu \nu}|0_S\rangle$, where $|0_B\rangle$ and $|0_S\rangle$ are the local vacuum states of the bulk and system respectively. The theory reduces to standard gravity when we have no field  excitations, i.e., $\rho_{S}=|0_S\rangle\langle0_S|$, the bulk is classical, i.e., $\langle \hat{T}^{B}_{\mu \nu}\rangle_{\rho_{B}}-\langle 0_B|\hat{T}^{B}_{\mu \nu}|0_B\rangle\sim T_{\mu \nu}^{B}$, and $T_{\mu \nu}^{B}$ well approximates a classical stress energy tensor. Furthermore, if the energy contained by the field can be considered a perturbation to the energy contribution of the bulk, we recover semiclassical gravity, since we have $\langle \hat{T}_{\mu \nu}\rangle_{\rho}-\langle \hat{T}_{\mu \nu}\rangle_{\rho_{p.}}\sim T_{\mu \nu}^{B}+\langle \hat{T}^{B}_{\mu \nu}\rangle_{\rho_{S}}-\langle 0_S|\hat{T}^{S}_{\mu \nu}|0_S\rangle$. This, together with the fact that we need to take averages following the prescription mentioned before, can provide a renormalised version of semiclassical gravity which can be employed as long as the field contribution is small compared to the contribution of the bulk.

If the state is separable and the theory non-interacting, the state $\rho_{B}$ is pure but $\rho_{S}$ is mixed, we then have that $\langle \hat{T}_{\mu \nu}\rangle_{\rho}-\langle \hat{T}_{\mu \nu}\rangle_{\rho_{p.}}=\langle \hat{T}^{B}_{\mu \nu}\rangle_{\rho_{B}}-\langle0_B| \hat{T}^{B}_{\mu \nu}|0_B\rangle+\langle \hat{T}^{S}_{\mu \nu}\rangle_{\rho_{S}}-\langle \hat{T}^{S}_{\mu \nu}\rangle_{p,S}$, where $\rho_{p,S}$ is the relevant passive state. An illuminating example is the case where $\rho_{S}$ is a thermal state. Then $\rho_{p,S}=\rho_{S}$ and our source would reduce to $\langle \hat{T}_{\mu \nu}\rangle_{\rho}-\langle \hat{T}_{\mu \nu}\rangle_{\rho_{p.}}=\langle \hat{T}^{B}_{\mu \nu}\rangle_{\rho_{B}}-\langle0_b| \hat{T}^{B}_{\mu \nu}|0_B\rangle$, which would imply that the system does not contribute to the gravitational field, albeit being excited. Note that, in this case, $\rho$ is a mixed state.

This conclusion is surprising. We would expect that an excited system, in a thermal state, would gravitate. However, we show below that the interplay between the purity of the global and reduced states determines the contributions to gravitation. This is one of our main results.

We now draw another major consequence of equation \eqref{quantum:einstein:equations} in the semicalssical picture, where bulk is classical and the field is initially in the vacuum. We note that both Hawking radiation in Schwarzschild spacetime and the Unruh effect \textit{do not occur}, i.e., there are no particles that can be detected by any observer, regardless of their motion, from the vacuum state. In fact, both spacetimes arise as solutions to Einstein equations in the vacuum with different boundary conditions \cite{BandD}. Therefore, any detection on behalf of a detector is purely a consequence of the coupling between the detector and existing field excitations, from which \textit{energy can be extracted}. In this sense, the Hawking and Unruh effect are not fundamental. This also implies, for example, that spherically symmetric \textit{static} black holes do not emit radiation and persist forever. 

It is easy to see that our semiclassical equations are consistent with Minkowski or Schwarzschild spacetimes, however, they predict that no excitations can be detected in the vacuum state of the field depending on the state of motion of the observer. We argue that, in light of all the considerations and motivations presented in this work, this conclusion should not be surprising. If the bulk, which provides the background metric, and the field do not interact or are not entangled, one should not expect any exotic phenomena to appear.

Interesting scenarios, such as black hole evaporation and gravitating thermal states, can be expected and can occur when the initial state is not separable or the theory includes interactions. For instance, let the state $\rho$ be pure but not separable in the bulk-system bipartition and the bulk and system not interact. In this case, the semiclassical regime of our main equation \eqref{quantum:einstein:equations} reads 
\begin{align*}
G_{\mu \nu}\sim\,T_{\mu \nu}^{B}+\langle \hat{T}^{S}_{\mu \nu}\rangle_{\rho_{red.,S}}-\langle 0_S|\hat{T}^{S}_{\mu \nu}|0_S\rangle.
\end{align*}
We notice that, in this case, the global vacuum of the theory $|0\rangle$ is still a product state of local vacua $|0\rangle=|0_B\rangle\otimes|0_S\rangle$, but the reduced state $\rho_{red.,f}$ of the system will in general be mixed. This assumption is reasonable in light of important results in entanglement theory that state that arbitrary subsystems of a system with large enough degrees of freedom are typically found in highly mixed states \cite{Popsecu:Short:2006}. Therefore, if the reduced state $\rho_{red.,f}$ is a thermal state, detection of field modes will also provide a thermal response and these particles will gravitate as expected. 

Finally, the initial state $\rho$ can be pure and not separable, within an interacting theory. In this case, we cannot decompose the total stress energy tensor $\hat{T}_{\mu \nu}$ as a sum of local contributions, and the complexity of the problem increases dramatically.

These consideration agree with the idea that vacuum-related processes, such as the Casimir force \cite{Genet:Intravia:2004} and particle creation via dynamical Casimir effect in a cavity \cite{Dodonov:2010}, are fundamentally connected to (time dependent) interactions of fields with matter within the boundaries \cite{Nicolic:2016}. Our work would also agree with the idea that radiation from black holes should be a consequence of the interaction of quantum fields with the (constituents of the) black hole itself.

When the field is in its vacuum state $|0_S\rangle$, there is no natural length scale in the theory. The theory predicts that the metric is the Minkowski metric and all expectation values vanish identically. The lack of a natural length scale in quantum field theory is an important signature of Lorentz invariance and leads to ultraviolet divergences \cite{Srednicki:2007}. Nevertheless, if the field is initially excited, the situation changes. Let us consider a situation where there is no bulk and the field is initially in a \textit{one particle} state $|1>$. The universal character of one particle states cannot be found self consistently in this theory but might be determined by a more fundamental theory. In order to have a well defined one particle state we require that the state $|1>$ be normalised, i.e., $<1|1>=1$. In this case, the particle's ``spatial extent'' cannot have support on the whole spacetime, as it occurs for standard plane wave solutions $u_{\boldsymbol{k}}\propto\exp[-i\omega\,t+\boldsymbol{k}\cdot\boldsymbol{x}]$ to the Klein-Gordon equation with sharp momenta $\boldsymbol{k}$, see \cite{BandD}. Instead, the particle will have a characteristic size $\sigma$, and will be mostly confined within a finite volume.

Given a (inverse) length scale $\sigma$ introduced by spatial extent of the excitations, we can then discuss important properties of our main equation \eqref{quantum:einstein:equations}. We normalise the coordinates $x^{\mu}$ by the $\sigma$, i.e., $x^{\mu}\rightarrow \sigma\,x^{\mu}$, and obtain
\begin{align} \label{quantum:einstein:equations:dimensionless}
G_{\mu \nu}=&\xi\,\left[\langle \hat{T}_{\mu \nu}^{B}\rangle_{\rho}-\langle0| \hat{T}_{\mu \nu}^{B}|0\rangle+\langle \hat{T}_{\mu \nu}^{S}\rangle_{\rho}-\langle0| \hat{T}_{\mu \nu}^{S}|0\rangle\right],
\end{align}
where we have assumed that the initial state is pure. The role of the dimensionless control parameter $\xi:=(l_P\,\sigma)^2$, where $l_P$ is Planck's length, has already been discussed in the context of gravitation with nonclassical states \cite{Bruschi:2016}.

We note that the vast majority of excitations and physical systems in the universe have a characteristic size $\sigma$ that satisfies $\xi\ll1$. This means that, for all purposes, the theory represented by \eqref{quantum:einstein:equations:dimensionless} is \textit{de facto} within a perturbative regime. One can then solve \eqref{quantum:einstein:equations:dimensionless} in a self consistent way by employing suitable perturbative approaches \cite{Bruschi:2016}. 

A few final comments are in place. i) The procedure to compute expectation values of physical quantities needs to be consistently changed in order to agree with \eqref{quantum:einstein:equations}. ii) Heisenberg equation $\rho(t)=U^\dag\,\rho(0)\,U$ might need to be modified to take into account the fact that bare energy is not the ``conjugate momentum'' of time. iii) The theory needs to be proven renormalisable. iv) The theory must predict the fate of the evolution of gravitating matter, such as collapsing matter to form black holes up to the Planck scale. v) Classical gravity is recovered, and is meaningful, only in the presence of ``enough'' energy and matter. Regardless of the success of direct applications of \eqref{quantum:einstein:equations}, we believe that the idea that only extractible work gravitates is fundamental and will be at the core of a successful theory of quantum gravity.

To summarise, we have introduced the idea that only extractible energy is a source of gravity. We have shown that the field and matter contribution to gravitation depends dramatically on the interplay between the purity of the global state (i.e., of the Universe) and the purity of the reduced states of the subsystems (i.e., fields and matter). We have discussed the first applications of our proposal and we have highlighted the most important questions to be answered. Our results suggest that small quantum systems gravitate differently than large ensembles of matter.
We believe that our results can help to understand better the overlap of relativity and quantum theories and, ultimately, in the quest of a unified theory of Nature.
We leave it to future work to develop this proposal to a full scale theory.


\textit{Acknowledgements} -- We thank Donato Giorgio Torrieri, Riccardo Rattazzi and Tommy Burch for useful discussions. We extend special thanks to Marcus Huber, for invaluable help with the aspects of thermodynamics in a quantum framework and to Jorma Louko, for clarifications of important issues of quantum field theory in flat and curved spacetime. D. E. B. acknowledges the Hebrew University of Jerusalem for hospitality, where part of this work was performed.

\bibliography{GravityWeightPaper}

\end{document}